\documentclass[12pt]{article}
\usepackage{latexsym,amssymb,epsf}

\begin{document}

\title{Higgs mechanism in the Randall-Sundrum model}

\author{A.~Flachi\thanks{e-mail:{\tt antonino.flachi@ncl.ac.uk}}
\ and David J.~Toms\thanks{e-mail:
{\tt d.j.toms@newcastle.ac.uk}}\\
Department of Physics, University of Newcastle upon Tyne,\\
Newcastle Upon Tyne, United Kingdom NE1 7RU}

\date{June 2000}
\maketitle

\begin{abstract}
We consider the dimensional reduction of a bulk scalar field in the Randall-Sundrum 
model. By allowing the scalar field to be non-minimally coupled to the spacetime 
curvature we show that it is possible to generate spontaneous symmetry breaking 
on the brane.
\end{abstract}
\eject

The idea that spacetime may have some extra dimensions, beyond the usual 
four of Einstein's theory, has been shown to provide an interesting solution to 
the gauge hierarchy problem~\cite{first}. An intriguing version of this scenario is the one 
proposed by Randall and Sundrum \cite{1} ---
a five dimensional model with one extra spatial dimension having an 
orbifold compactification. Essentially the model consists of two three-branes 
with opposite tensions sitting at the two orbifold fixed points. 
The 5-dimensional line element is 
\begin{equation}
	ds^2 = e^{-2kr|\phi|} \eta_{\mu \nu}  dx^{\mu} dx^{\nu} - r_c^2 d\phi^2~.
		\label{rsmetric}
\end{equation}
where $x^{\mu}$ are the 4-dimensional coordinates, 
and $|\phi|\le \pi$ with the points $(x^{\mu},\phi)$ and $(x^{\mu},-\phi)$ 
identified. The factor of $ e^{-2kr|\phi|}$ present in (\ref{rsmetric}) 
means that the Randall-Sundrum spacetime is not a direct product of 4-dimensional spacetime and 
the extra fifth dimension. This factor is often referred to as the warp factor.
The three-branes sit at $\phi = 0$ and $\phi= \pi$. $k$ is a constant of order of the 
Planck scale and $r_c$ is an arbitrary constant associated with 
the size of the extra dimension. 
The interesting feature of this model is the simple 
way in which it generates a TeV mass scale from higher dimensional Planck 
scale quantities.
A field with mass $m_0$ which is confined on the negative tension brane 
will develop a physical mass $m=m_0e^{-kr_c\pi}$; therefore
the electroweak scale is naturally realized if one adjusts the length of the 
extra-dimension to $kr_c\sim 12$. In fact, in the original version of the 
Randall-Sundrum model all the standard model particles are supposed to 
be confined on the brane with only gravity living in the bulk (5-dimensional) 
spacetime.

An attractive alternative to confining standard model particles on the brane is to 
allow all of the fields to live in the bulk spacetime.
Several aspects of this model have been studied and the literature is 
already immense. In \cite{2}, the physics of a bulk scalar field is 
studied, and the authors point out that the warp factor in (\ref{rsmetric}) localises the field 
on the brane. This situation has been further explored in \cite{3,4}, 
where bulk gauge bosons have been considered. 
Some aspects of fermion bulk fields have been explored in \cite{17,18,19,gh}, 
and bulk supersymmetry has been studied in \cite{gh}.
Both fermions and gauge bosons in the bulk have been studied in \cite{5,riz}. 
In this last case 
it turns out that the zero modes, interpreted as standard model particles, 
are localised on the brane, explaining why the 
hierarchy problem is solved in a different setting with respect to the original 
Randall-Sundrum model in which the standard model was confined on the wall. 
In \cite{gh,riz,21,6} a bulk Higgs field as the origin of spontaneous
symmetry breaking was also 
considered. Unfortunately phenomenological constraints for gauge 
boson masses to be of the order of the electroweak scale requires a hierarchically 
small Higgs mass. This seems to rule out a bulk Higgs field, leaving two alternatives: 
stick the Higgs on the brane by hand (or by some as yet unknown confinement mechanism), 
or expect some dynamics to drive a bulk scalar field to a negative mass squared 
field in four dimensions. Other arguments have been given in \cite{riz,21}.

In this letter we investigate the possibility of obtaining spontaneous symmetry 
breaking as a consequence of dimensional reduction within the Rand\-all-Sun\-drum 
model with fields living in the bulk. We will show how it is possible to obtain a scalar 
particle with imaginary mass via a Kaluza-Klein reduction in the Randall-Sundrum spacetime. 
A scalar field in a higher dimensional spacetime has been shown to 
reduce in four-dimensions to 
the Kaluza-Klein infinite tower of scalar fields whose 
masses $m_n$ are quantised \cite{2}. 
For the Randall-Sundrum metric the masses $m_n$ are given by solutions 
of the transcendental equation
\begin{equation}
	0=y_\nu (ax_n)j_\nu(x_n) - j_\nu (ax_n)y_\nu(x_n)\;,
	\label{spectrumgw}
\end{equation}
where we have defined $a=e^{-\pi k r_c}$ and $m_n=kax_n$, with $x_n$ the 
$n^{th}$ positive solution to (\ref{spectrumgw}). The functions $j_\nu$
and 
$y_\nu$ are given by the following combinations of Bessel functions:
\begin{eqnarray}
	j_\nu(z) &= &2J_\nu(z) + zJ_{\nu}^{\,\prime}(z)\;,
	\label{littlej}\\
	y_\nu(z) &=& 2Y_\nu(z) + zY_{\nu}^{\,\prime}(z)\;.
	\label{littley}
\end{eqnarray}
The order of the Bessel functions is $\nu = \sqrt{4+{\hat{m}^2 \over k^2}}$ 
where $\hat{m}$ is the mass of the five-dimensional scalar field.

Our model is described by a simple generalisation of that in \cite{2}~:
\begin{equation}
	S={1\over 2} \int d^4x \int_\pi^\pi d\phi \sqrt{g}
	\left( g^{AB} \partial_A \Phi \partial_B \Phi - \hat{m}^2 \Phi^2 - \xi 
	\hat{R} \Phi^2 - {\lambda \over k} \Phi^4 \right)\;.
	\label{action}
\end{equation}
Here $\lambda$ is dimensionless (for the moment no value for it is 
specified) and $\xi$ is also dimensionless and represents a non-minimal 
coupling of the scalar field to the gravitational background. $g_{AB}$ represents 
the metric of (\ref{rsmetric}) and $\hat{R}$ is the scalar curvature computed from 
this metric. We will see that it is the non-minimal coupling of the scalar field which 
allows the possible generation of Higgs particles in the theory.

Let us now turn to Kaluza-Klein reduction. In the non-minimally coupled case the 
situation is different from the one presented in \cite{2} due to the 
presence of the $\hat{R}\phi^2$ term. Decomposing the fields as a sum over 
modes and setting $\lambda=0$ initially (as in \cite{2}), we write
\begin{equation}
	\Phi(x,y) = \sum_n \psi_n(x) f_n(y)\;,
\end{equation}
with 
\begin{equation}
	\int_{\pi r_c}^{\pi r_c} dy e^{-2\sigma} f_n(y)f_n(y) = \delta_{nm}\;.
	\label{ort}
\end{equation}
The field equation for the modes becomes
\begin{eqnarray}
	-e^{2\sigma} \partial_y (e^{4\sigma} \partial_y f_n (y) ) + m^2 e^{-2\sigma} 
	f_n (y) &&\nonumber\\
- 16 k \xi e^{-2\sigma} \left( \delta(y) - \delta(y-\pi r_c)\right)
	f_n(y)&&
=m_n^2 f_n(y)\;,\label{modeq}
\end{eqnarray}
where we have defined $m^2 = \hat{m}^2 +20 \xi k^2$. It can be seen that the presence of the terms 
involving Dirac delta distributions in (\ref{modeq}) has its origin in the curvature of the Randall-Sundrum spacetime.
For $y\ne0$ or $y\ne\pi r_c$ the linearly independent solutions to (\ref{modeq}) are the same 
Bessel functions as those given 
in \cite{2}. After applying the boundary conditions appropriate to the orbifold compactification 
in the model it is easily shown that
\begin{equation}
f_n(y)= N_n e^{2\sigma} \left( J_\nu\left({m_n \over k}e^{\sigma}\right) - 
\left({j_\nu({m_n \over k})\over y_\nu({m_n \over k})} \right) 
Y_\nu\left({m_n \over k}e^{\sigma}\right) \right)
	\label{modes}
\end{equation}
with $\nu = \sqrt{4 + {m^2 \over k^2}}$ and 
\begin{equation}
	j_\nu(z) = (2+8\xi)J_\nu(z) + zJ_{\nu}^{\,\prime}(z)~,
	\label{littlejxi}
\end{equation}
\begin{equation}
	y_\nu(z) = (2+8\xi)Y_\nu(z) + zY_{\nu}^{\,\prime}(z)~.
	\label{littleyxi}
\end{equation}
For $\xi=0$ these results reduce exactly to those in \cite{2}, as they should.
The normalization factor can be evaluated in closed form, but the expression 
is very lengthy and will not be given explicitly here.

Let us now turn to the mass spectrum. As was said before, it is interesting to 
extend the Randall-Sundrum model by considering the possibility of having other bulk 
fields. 
Since spontaneous symmetry breaking in the bulk appears to be disfavoured for a 
variety of reasons, the Higgs field is forced to live on the brane. 
No alternative to generate spontaneous symmetry 
breaking on the brane starting from an ordinary bulk scalar field has been 
investigated. 
We want to show that a non-minimally coupled scalar field offers such a 
possibility.

The masses of the Kaluza-Klein excitations are given by the zeroes of the 
function
\begin{equation}
    F_\nu(z, \xi) = y_\nu (az)j_\nu(z) - j_\nu (az)y_\nu(z)~.
	\label{spectrum}
\end{equation}
Clearly, the previous considerations lead us to look for a Higgs field 
after Kaluza-Klein reduction, which in turn means looking for purely imaginary 
solutions of (\ref{spectrum}).
After rotating to the complex plane ($z \rightarrow iz$), $F_\nu(z, \xi)$ 
can be re-written in terms of the modified Bessel functions and the mass 
spectrum equation reads
\begin{equation}
    0 = i_\nu (az)k_\nu(z) - k_\nu (az)i_\nu(z)~.
	\label{modified}
\end{equation}
Obviously the previous equation cannot be solved analytically and its 
numerical study is rather tricky because of the oscillating behaviour of 
the Bessel functions, the presence of an extremely small 
exponential factor and of the fact that $F_\nu(z, \xi)$ becomes very
large. 

Although it is necessary to resort to a numerical analysis, several points can be 
addressed analytically. First of all, note that the function (\ref{modified}) does not admit real 
zeros unless the order of the Bessel function is imaginary. 
This can be achieved by letting $\xi<-{1\over 20}(4+{m^2\over k^2})$. 
Secondly, although the modified Bessel functions appearing in the above 
transcendental equation are complex, the combination appearing in (\ref{modified}) can 
be shown to be real. Thus it is possible to expedite the numerical procedure by taking
the real part.

A negative value of $\xi$ is necessary if we are to obtain a Higgs type mass for the 
dimensionally reduced field. The coupling constant $\xi$ is to be regarded as a free parameter. 
Although a popular choice is to fix its value to zero or the conformal value 
($1/6$ for 4-dimensions, and $3/12$ for 5-dimensions) for 
computational simplicity, there is no good argument to prefer any specific value.
Indeed, there are specific theories in which a prescription for $\xi$ does exist. 
For instance, it has been shown in \cite{6} that Higgs scalar fields must have 
$\xi \leq 0$ or $\xi \geq 1/6$ in order to have an absolutely stable ground 
state. 
Other examples have been considered in \cite{7,8,9}.

In performing the numerical analysis various features have to be taken into account.
The first is related to the bounds on the Higgs mass. 
The discussion on this issue is quite complicated because in the Randall-Sundrum
model, higher dimensional operators need to be included in order to have a 
reliable upper limit on the Higgs mass \cite{datta}. 
However, the mass spectrum does not depend dramatically on the value of 
the Higgs mass--- this can be easily seen numerically--- therefore we do not expect 
our analysis to change drastically by altering the Higgs mass within a reasonable 
range of values. 
It is interesting to note that from the knowledge of the bounds on the Higgs 
mass, it is possible to trace back bounds on $\xi$. In the following we
fix $m_H= 100$~GeV.

\begin{figure}
\begin{center}
\leavevmode
\epsfxsize=0.75\textwidth
\epsffile{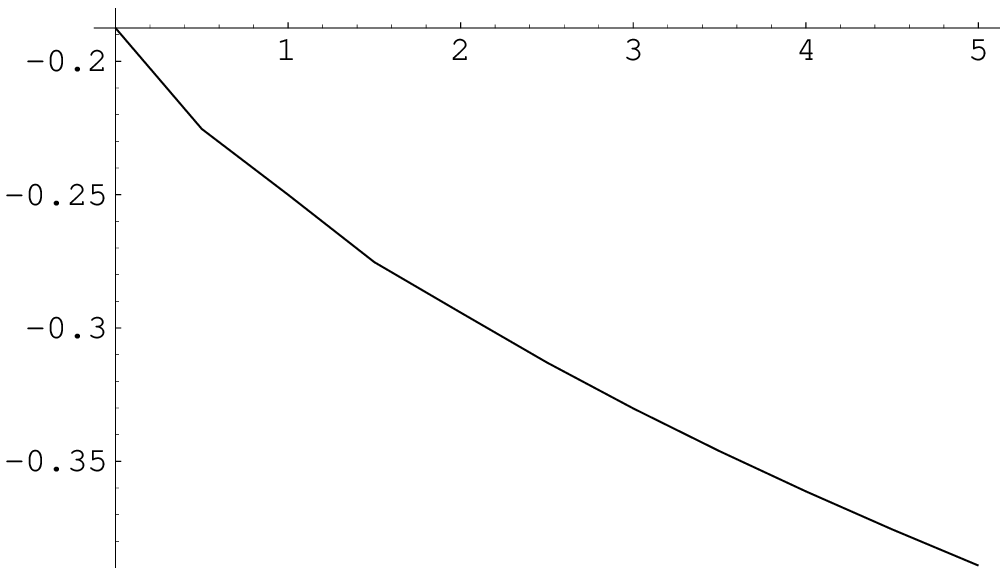}
\end{center}
\caption{\small Plot of $\xi$ versus $\left({\hat{m}\over k}\right)^2$.}
\label{fig}
\end{figure}

Another parameter, which seems to be quite problematic, is the ratio 
${\hat{m}\over k}$, or simply $\hat{m}$, which needed to be finely 
tuned when placing the Higgs field in the bulk in previous work\cite{6}. 
We have chosen ${\hat{m}\over k}=1$, again noting that $F_\nu(z,\xi)$ 
is not particularly sensitive to this parameter for values between $0$ and $5$. The 
dependence of $\xi$ on ${\hat{m}\over k}$ has been studied and the 
results are shown in (fig \ref{fig}).

The approximate value of $\xi$ corresponding to a mass of 100 GeV is
$\xi=-0.250020221625119498$. The standard model bounds \cite{sm1,sm2} on the 
Higgs mass correspond to $\xi$ varying between $-0.250019$ and $-0.250061$. 
It is important to note that for a fixed value of the Higgs mass $\xi$ can 
assume different values. In our computation we took the first zero. As we will discuss,
this has the virtue that the next mass eigenvalue (if there is one) is extremely large.

When mode expanding the 5-dimensional action we have to consider 
the self interaction term, which simply needs to be integrated over the 
extra co-ordinate. Taking into account only the first low lying mode, after 
integration, the self interaction term looks like 
\begin{equation}
	S_{int} = \lambda_{eff} \int d^4x \psi_n^4(x)~,
\end{equation}
with 
\begin{equation}
	 \lambda_{eff}= {2\lambda \over k} \int_0^{\pi r_c} dy e^{-4ky} f_n^4(y)~.
\end{equation}
Although it does not appear possible to obtain an exact closed form result 
for this expression, it can be evaluated numerically without difficulties and 
the result is $\lambda_{eff}=9.42553~10^{-18}~\lambda$. Even for very large 
values of $\lambda$ in the $5$-dimensional theory, the self-interaction term in 
the effective $4$-dimensional theory is small.

An important feature to consider is the presence of higher Kaluza-Klein modes. 
It is important to make sure that there are no other low lying modes for at 
least two reasons: firstly, because their presence would considerably complicate 
any phenomenological analysis; secondly, because they would 
give rise in the dimensional reduction of the self-interaction portion to 
mixed terms. 
Therefore we had to extend our numerical investigation not only to the 
first mode, but to higher modes as well, to check the reliability of the 
present model. 
Because of numerical limitations, we have studied the 
function $F_\nu(z, \xi)$ only in a relatively large region, $|z|<740000000$ 
corresponding to modes of masses $m<10^{11}~GeV$. Our numerical study 
shows that there is only one root in this region, enabling us to discard in 
the Kaluza-Klein expansion for the self interaction term the mixed modes 
safely if we are only interested in the low energy effective theory.

In conclusion, we have discussed the issue of a bulk Higgs field, relevant 
in any attempt to place the standard model in the bulk. Since this possibility 
is disfavoured, we have indicated a mechanism which admits a bulk scalar 
field and spontaneous symmetry breaking on the brane. In other words we have 
considered a bulk scalar field with a positive mass term and indicated a way of 
obtaining a scalar field with imaginary mass on the brane. We achieved this 
by considering a non-minimally coupled theory and letting the scalar coupling  
be negative. The non-minimal coupling constant is responsible for this. 
Thus it is possible to generate spontaneous symmetry breaking in a natural 
way using the geometry.

A.~Flachi would like to thank the University of Newcastle upon Tyne for the award 
of a Ridley Studentship.

\end{document}